\documentclass{aa}
\begin{document}

   \thesaurus{20     
              (02.12.2;  
               03.20.8;  
               04.03.1;  
               06.16.2)} 
   \title{A catalogue of  accurate wavelengths in the optical spectrum of the Sun}

\author{C.~Allende Prieto \and R. J.~Garc\'{\i}a L\'opez}

\offprints{C.~Allende Prieto (callende@iac.es)}

\institute{Instituto de Astrof\'\i sica de Canarias, E-38200 La Laguna, 
           Tenerife, Spain}                     
  
\date{Received date, ; accepted date, }

   \maketitle

	\markboth{Allende Prieto \& Garc\'{\i}a L\'opez: Accurate Solar Wavelengths}{}

   \begin{abstract}

 We present accurate measurements of the central wavelengths of 4947
atomic absorption lines in the solar optical spectrum. The wavelengths,
precise to a level $\sim 50-150$ m s$^{-1}$, are given for both flux
and disc-centre spectra, as measured in relatively recent FTS solar
atlases. This catalogue modernizes existing sources based on
photographic measurements and provides a benchmark to test and perform
wavelength calibrations of astronomical spectra. It will also permit
observers to improve the absolute wavelength calibration of solar
optical spectra when lamps are not available at the telescope.


\end{abstract} 

%

\section{Introduction}
\label{sec1}

Wavelength calibration is almost always needed in the process of
producing useful astronomical spectra. To calibrate accurately is a
non-trivial problem, in particular when working at high or very-high
spectral resolution. Fourier transform spectrographs (FTS) are
specially well-suited to this task, but they are not readily applied in
conditions requiring high spatial or time resolution, so grating
spectrometers are much more commonly used for astronomical
observations. In this case, it becomes necessary to set reference
positions corresponding to known wavelengths on the detector.  This can
be achieved by using very sharp observed telluric lines, but their
location in the spectrum cannot be chosen by the astronomer. It is very
usual to find spectral calibration lamps available for use with an
astronomical spectrograph. The emission lines produced in the lamps
have been previously measured at the laboratory, and this method
usually provides a valid reference frame. However, it is often
impractical to expose the calibration lamp simultaneously with the
astronomical target and, unless the spectrograph is installed at a very
stable focal station, the position of the spectrum on the  detector
varies depending on the telescope position. Accuracy is then limited by
the instrument characteristics and observations of the calibration
lamps are required between successive astronomical exposures.
Nonetheless, calibrations via arc or hollow cathode spectra are
normally accurate enough for most purposes.

Ingenious techniques have been used to improve the accuracy of wavelength
calibrations, such as placing gas cells at the entrance of the
spectrograph (e.g., Deming \& Plymate 1994), but it is rare to find such
systems available and convenient for regular observations.

On occasion the available lamps are not very rich in lines in the
spectral range of observation. In some circumstances, an external check
of the final precision in the translation into wavelengths would be
desirable. One method for tackling problems such as these is to use
solar spectra as templates.  Changes in the wavelengths of the lines in
the  integrated sunlight spectrum around the solar cycle have been
proved to be very small, bellow some 15 m s$^{-1}$ (Jim\'enez et al.
1980; Wallace et al. 1988; McMillan et al. 1993; Deming \& Plymate
1994).  At 5000 \AA, this translates into $\sim$ 0.3 m\AA, so the solar
spectrum does offer a  very stable source. In most practical cases, the
accuracy will be imposed by the spectral resolution achieved. During
night-time observations, the solar flux spectrum is observable after
reflection from the Moon.

Measurements of solar wavelengths in the integrated solar optical
spectrum were published in 1929 by Burns and collaborators (Burns 1929;
Burns \& Kiess 1929; Burns \& Meggers 1929), using photographic
detectors and a grating spectrograph. The relatively recent solar flux
FTS atlases offer a much higher quality spectrum of the Sun
seen as a star.

As the solar spectrum is so intense, on some solar telescopes no
calibration lamps are deemed necessary, and the wavelength scale is set
using the solar spectrum itself. Reasonable precision can be reached
using the spectrum at the centre of the disc to compare with previously
measured disc centre wavelengths, thus avoiding differential shifts due
to the limb effect. In this case, small scale motions have to be
averaged out, integrating in time and/or space, in order to minimize
errors.

The {\it Kitt Peak Table of Photographic Solar Spectrum Wavelengths}
(Pierce \& Breckinridge 1973) has been extensively used by solar
observers to set up the wavelength scale on their spectra. These
observations, made on photographic plates, have been superseded in
quality by the more recent FTS observations at the centre of the disc.

To improve on the various sets of photographically based measurements
(which date back to 1930 in the case of the solar flux spectrum),
provide them in a homogeneous machine-readable format, use them to test
spectral calibrations of very high resolution stellar spectra (e.g.,
Allende Prieto et al. 1995), and improve the accuracy of our own solar
observations, we have determined the position of the central
wavelengths of 4947 atomic lines in the optical solar spectrum. The
employed source solar atlases, prepared from FTS data, and the
fashion in which we performed the measurements is described in the
succeeding sections.

\section{The source atlases}

Among other solar flux FTS atlases, the {\it Solar Flux Atlas from
296 to 1300 nm} (Kurucz et al. 1984), which is available from the
NOAO\footnote{National Optical Astronomical Observatories, USA} ftp
site, provided us with a high-quality spectrum of the Sun seen as a
star. It was obtained at the McMath\footnote{Nowadays renamed the
McMath/Pierce telescope} telescope at Kitt Peak.

The FTS disc-centre spectrum included in the newer {\it Spectral Atlas
of Solar Absolute Disk-Averaged and Disk-Center Intensity from 3290 to
12510 \AA} (Brault \& Neckel 1987; for details see Neckel 1994) was
also obtained at the McMath telescope. The wavelengths of selected
lines from the table of Pierce \& Breckinridge (1973), which was
produced from observations with a grating spectrometer and a
photographic detector at the same telescope, have been the base for
placing its wavelength calibration on an absolute scale. We have
measured line central wavelengths in this atlas, available to us as
part of the IDL KIS\footnote{Kiepenheuer-Institut f\"ur Sonnenphysik,
Freiburg, Germany} library. A flux spectrum is also contained in this
FTS atlas, which shares the source data obtained by J. Brault and
collaborators with the atlas prepared by Kurucz et al. (1984).

The atlases cited achieve signal-to-noise ratios of about 2500 and a
resolving power $\lambda /\Delta\lambda \sim$ 400000. A quantitative
basis for confidence in these atlases has been established by comparison
between central wavelengths of 1446 Fe\,{\sc i} lines in the solar
spectrum and at rest, performed by Allende Prieto \& Garc\'{\i}a L\'opez
(1998).  Briefly, they found:

\begin{itemize}

\item The atlases have been corrected for all Doppler shifts between the
 centre of mass of the Sun and Earth, and the maximum shift to the red
exhibited by the lines studied is the gravitational shift: 636 m
s$^{-1}$, corresponding to a null convective blue shift.

\item There is no stronger than expected trend of the line shifts with
wavelength. 

\item  A clean correlation exists between the equivalent width and the
line shift, reaching a plateau near the gravitational redshift for
lines stronger than 200 m\AA.

\end{itemize}

\section{The Catalogue}

This work enlarges the measurements of Allende Prieto \& Garc\'{\i}a
L\'opez (1998) to many other species, with the aim of optimizing its
use for wavelength calibration. The line list from Th\`evenin (1989,
1990), including 6606 lines   classified by Moore et al. (1966) as
singly blended or unblended, has been chosen as a guide to select the
features to be measured. 

A fourth-order polynomial was fitted to the 50 m\AA\ wavelength
interval around the line minimum to find the line centre as precisely
as possible. In the electronic version of the atlases, this
corresponds to 11 points for the  solar flux spectrum and 25 points
for the centre-of-the-disc spectrum. Errors in the wavelength
determination were estimated by translating the standard deviation of
the fit into the corresponding units of the wavelength axis,
neglecting the extremely low photometric noise.

Table 1, available only in electronic form from the CDS, lists the
central wavelengths in the flux and disc-centre solar spectra, the
errors of the measurements, and the element identification, excitation
potential and 'solar' oscillator strengths ($\log gf$; from Th\`evenin
1989, 1990) for 4947 lines between 3944 and 7960 \AA\ included in
Th\`evenin's list. More than one line corresponding to the same feature
is listed when the identification is not clear from the wavelengths
published by Th\`evenin.

A comparison between a limited sample of wavelengths (42 lines) listed
by Pierce \& Breckinridge (1973) and those we have measured in the
atlas of the disc-centre shows that the absolute scales agree
(difference: 79 $\pm$ 6 $\times$ 10$^{-5}$ \AA) and that the relative
differences are fully accounted for by a straightforward consideration
of the expected errors in the photographic atlas ($\sim$ 2.5 m\AA). 

Errors of the wavelengths measured in the flux spectrum can be as
large as 100 m s$^{-1}$, as quoted by Kurucz et al. (1984). For the
wavelengths in the spectrum of the disc centre, systematic errors are
given from the comparison with the photographic atlas of Pierce \&
Breckinridge (1973), which has been used as reference. They estimate
the absolute accuracy from comparison with the interferometrically
determined wavelengths of Adam (1952, 1958) and Nichols \& Clube
(1958), arriving at 0.3 m\AA. Our systematic errors are then of the
order of $\sqrt{0.8 ^{2} + 0.3 ^{2}} \sim $ 0.9 m\AA, while FTS
intrinsic errors are about 10 m s$^{-1}$ (Neckel \& Labs 1990),
yielding a final precision of about 50 m s$^{-1}$ at $\lambda$ 5000
\AA. These numbers allow us to claim that errors quoted in Table 1
(typically $\sim 50-150$ m s$^{-1}$) are conservative.

The present database updates others which were previously available
based on photographic spectra, and the catalogue is presented in an
homogeneous machine-readable format available via internet from the
CDS. The wavelengths measured conform a reference frame which can be
used to calibrate or field test calibrations of astronomical  spectra
by comparing with the solar spectrum, and to set up an accurate
wavelength scale for solar spectra, among other applications.

\begin{acknowledgements}

 We thank H. Neckel for his help in dealing with his solar atlas. We
wish to extend our gratitude to F. Th\'evenin, who has kindly lent a
digital copy of his line list, M. Collados for fruitful discussions on
the wavelength calibration of solar spectra, J. E.  Beckman for careful
reading of the draft and H. H. R. Kroll for performing the installation
and maintenance of the KIS computer libraries at the IAC.  NSO/Kitt
Peak FTS data used here were produced by NSF/NOAO.

This work was partially supported by the Spanish DGES under projects
PB92-0434-C02-01 and PB95-1132-C02-01.

\end{acknowledgements}

\end{document}